# Core crystallisation in evolving white dwarf stars from a pile-up in the cooling sequence

Pier-Emmanuel Tremblay[1,*], Gilles Fontaine[2], Nicola Pietro Gentile Fusillo[1], Bart H. Dunlap[3],

Boris T. Gänsicke[1], Mark Hollands[1], J. J. Hermes[3,4], Thomas R. Marsh[1], Elena Cukanovaite[1]

& Tim Cunningham[1]

[1] Department of Physics, University of Warwick, Coventry CV4 7AL, UK

[2] Département de Physique, Université de Montréal, C. P. 6128, Succursale Centre-Ville, Montréal, QC H3C 3J7, Canada

[3] Department of Physics and Astronomy, University of North Carolina, Chapel Hill, NC 27599, USA

[4] Hubble Fellow

**White dwarfs are stellar embers depleted of nuclear energy sources that predictably cool over billions of years at the expense of their leaking thermal reservoir of ions[1]. These stars, supported by electron degeneracy pressure, are essential to develop our understanding of dense plasmas, reaching densities of $10^7$ g/cm$^3$ in their cores[2]. It is predicted that a first-order phase transition occurs during white dwarf cooling, leading to the crystallisation of the non-degenerate carbon and oxygen ions in the core that releases a significant amount of latent heat and delays the cooling process by about one billion years[3]. Here we report the presence of a pile-up in the cooling sequence of white dwarfs within 100 pc of the Sun, using photometry and parallaxes determined from the *Gaia* satellite[4]. Based upon modelling, we infer that the pile-up arises from the release of latent heat as the cores of the white dwarfs crystallise. In addition to the release of latent heat we find strong evidence that cooling is**

**further slowed by the liberation of gravitational energy from element sedimentation in the crystallising cores[5-7]. Our results demonstrate the total energy released by crystallisation in strongly coupled Coulomb plasmas[8-9], and the newly measured cooling delays could improve the accuracy of using white dwarfs for age-dating stellar populations[10].**

The white dwarf cooling age at which crystallisation sets in is predicted to depend on the mass (Fig. 1), with more massive white dwarfs entering this phase transition earlier[3]. Another major event in the evolution of a white dwarf is the direct coupling between the degenerate core and the convective envelope[11], resulting in an initial slow down in cooling rates followed by an increase. At the low white dwarf masses (~0.55 $M_\odot$) of the old stellar populations in globular clusters, this event occurs at a similar age as crystallisation but has a stronger signature[12]. Previous attempts to measure cooling effects from crystallisation in globular clusters have therefore provided indirect evidence, based on linking the white dwarf and turn-off age determinations[13]. In contrast, crystallisation occurs much earlier than convective coupling in white dwarfs more massive than 0.7 $M_\odot$. The observational implication of this, predicted over fifty years ago[3], is an isolated crystallisation sequence in the colour versus absolute magnitude Hertzsprung-Russell (H-R) diagram, yet no direct observational evidence existed until now to characterise this event.

Because of their small radii, typically on the order of 0.01 $R_\odot$, white dwarfs are intrinsically faint, and consequently until recently very few had accurate distance measurements needed to measure their luminosities[14]. The second Data Release of the European Space Agency *Gaia* mission[4] (*Gaia* DR2) has led to a breakthrough, defining the first empirical cooling sequence of field white dwarfs in the H-R diagram[15]. While previous studies have investigated the cooling sequences of white dwarfs in old globular clusters[16], only a local volume-limited sample will contain white dwarfs spanning the full ranges of total ages and initial masses[17].

We used a recently established catalogue of high-confidence *Gaia* white dwarf candidates[18] to extract degenerate stars within 100 pc. The selection function of *Gaia* was found to be colour and magnitude independent down to the sky-position dependent faint magnitude limit[18] and the median parallax precision of 1.5 percent allows for unambiguous transformation to distances. For 15,109 sources the *Gaia* photometry and astrometry are reliable enough to derive the surface temperature, surface gravity, and mass[18] by fitting the data to model atmospheres[19] and standard evolutionary tracks with $^{12}$C/$^{16}$O core-composition and thick H envelopes[11]. Sloan Digital Sky Survey (SDSS) spectroscopy is available for 1309 of the white dwarfs within 100pc, providing their atmospheric composition.

The 100 pc field white dwarf cooling sequence exhibits a substantial amount of structure (Fig. 2). The bifurcation into two sub-sequences in the range $-0.1 < G_{BP} - G_{RP} < 0.6$ has been shown to be a split between H atmospheres in the upper branch and He-dominated atmospheres in the lower branch[18,20]. These two branches correspond to cooling tracks at the median (~0.6 $M_\odot$) white dwarf mass. More massive white dwarfs have larger absolute magnitudes because of their mass-radius relation[21] and are therefore expected to populate the area below the principal branches. A third separate "transverse" sequence is visible at fainter absolute magnitudes. Unlike the aforementioned bifurcation, this transverse feature is inconsistent with a single-mass cooling track. This rules out a simple astrophysical explanation such as effects from the mass loss in post-main-sequence evolution or merger products from binary evolution, as these scenarios cannot conceivably result in a tight white dwarf mass versus surface temperature correlation unrelated to the cooling process.

The transverse sequence fully coincides with the range of absolute magnitudes and colours at which the bulk of the latent heat from crystallisation is released for white dwarfs over the full range of masses. The crystallisation sequence is more clearly visible when the sample is restricted to white dwarfs with more simple hydrogen-dominated atmospheres, and for which independent spectroscopic parameters determined from fitting the hydrogen lines[19,22,23] agree with their position in the H-R diagram (Fig. 3). Roughly eight per cent of sources within the crystallised sequence harbour large (>2 MG) global magnetic fields detected from Zeeman splitting[18]. Helium-atmosphere white dwarfs also populate the cooler and less massive (< 0.7 $M_\odot$) area of the sequence. There is a dearth of massive helium-atmosphere stellar remnants in all parts of the H-R diagram including the crystallised sequence, which is likely caused by single star evolution not forming thin-hydrogen layers for higher mass progenitors[24]. The 100 pc sample was cross-matched with the *Galex*, 2MASS, *WISE*, Pan-STARRS and SDSS photometric data sets, and it was determined that white dwarfs within the transverse sequence are under luminous at all wavelengths compared to objects on the dominant cooling sequence, and therefore behave as genuine high-mass objects. We conclude that nothing stands out in the atmospheric properties of the white dwarfs in the crystallised sequence, apart from a tight correlation between colour and absolute magnitude. The consistent explanation is crystallisation, a cooling effect that is expected to impact white dwarfs of similar mass and interior composition at the same age, with little influence from the atmospheric composition or the presence of magnetic fields[2,25].

The crystallised sequence is not a cooling track but a mass-dependent pile-up across the H-R diagram resulting from the white dwarfs spending more time at this location as they release their latent heat. To further characterise this process we have extracted the white dwarf luminosity function in the mass range 0.9-1.1 $M_\odot$ from the *Gaia* 100 pc sample (Fig. 4). Two peaks are clearly seen in the luminosity function, one at higher luminosities that is attributed to crystallisation, and the other one at lower luminosities which is unambiguously linked to the finite age of the Galactic disk[10]. At lower masses than those considered, crystallisation occurs at fainter absolute magnitudes where it overlaps both with the convective coupling of the core with the envelope and the peak in the luminosity function caused by the age of the Galactic disk.

We have performed white dwarf population simulations (Fig. 4) assuming constant stellar formation over the past 10 Gyr, the Salpeter initial-mass function, a standard initial-to-final mass relation[26] coupled with predicted main-sequence lifetimes[27], and a *Gaia* magnitude limit of $G = 20$. These input parameters do not influence the slope of the luminosity function where crystallisation occurs, so we made no attempt to fit them to the observations. In contrast, the three simulations presented in Fig. 4 use different assumptions about the crystallisation process, showing a strong influence on the prediction of a peak at $-3.75 < \log L/L_\odot < -2.75$. The case without latent heat release by crystallisation is clearly ruled out by the observations. When latent heat is included in the modelling, there is, comparatively, a substantial increase of the predicted number of white dwarfs in the range of luminosity of the observed peak. The *Gaia* luminosity function is best reproduced when $^{16}$O sedimentation is allowed to occur along with the release of latent heat. Compared to the original $^{12}$C/$^{16}$O fluid mixture, sedimentation leaves behind a solid region that is oxygen-enriched, the level of which depends upon the actual composition of the fluid. The extra carbon in the fluid phase is forced upward from a crystallising shell, leading to a release of potential energy that further delays the cooling[7]. Note that this third model, with latent heat and phase separation, provides an excellent description of the overall *Gaia* luminosity function, including its descending branch. The cumulative cooling delay from crystallisation has a direct effect on the descending branch; stars that have more internal energy become warmer for the same age. The model bump is not perfectly modelled, as the observed feature is narrower and of higher magnitude, but its exact shape depends on several choices including the value of the Coulomb plasma parameter[28] ($\Gamma = 175$ here), the assumed chemical profile in the core ($^{12}$C/$^{16}$O in 50/50 proportions by mass and distributed homogeneously), the envelope stratification ($M_H/M_{WD} = 10^{-4}$ and $M_{He}/M_{WD} = 10^{-2}$), as well as possible $^{22}$Ne sedimentation[9,13]. More fundamentally, the existence and location of this bump provides the extremely strong evidence that the observed excess of white dwarfs in the *Gaia* transverse sequence bears the signature of crystallisation.

We report direct evidence that a first-order phase transition really occurs in high-density Coulomb plasmas[3], a theory that cannot be tested in laboratories because of the extreme densities involved, thus providing strong constraints on dense plasma physics[7-9,28]. Crystallisation significantly slows down the cooling process in white dwarfs and the observations also require the release of gravitational energy from the separation of an initially homogeneous fluid into a stratified solid with $^{16}$O/$^{12}$C ratio that increases towards the centre of the star, providing a new method to test nucleosynthesis processes in low and intermediate-mass stars[29]. The descending branch of the empirical white dwarf luminosity function is heavily impacted by phase separation[5-7] and quantum effects in Debye cooling[6,30], necessitating the understanding of these processes when relying on stellar remnants for age-dating stellar populations[10,11].

References

[1] Mestel, L. On the theory of white dwarf stars. I. The energy sources of white dwarfs. *Mon. Not. R. Astron. Soc.* **112**, 583-597 (1952).

[2] Tassoul, M., Fontaine, G. & Winget, D. E. Evolutionary models for pulsation studies of white dwarfs. *Astrophys. J. Supp.* **72**, 335-386 (2010).

[3] van Horn, H. M. Crystallization of White Dwarfs. *Astrophys. J.* **151**, 227-238 (1968).

[4] Gaia Collaboration *et al*. Gaia Data Release 2. Summary of the contents and survey properties. *Astron. Astrophys.* **616**, A1 (2018).

[5] Garcia-Berro, E., Hernanz, M., Mochkovitch, R., & Isern, J. Theoretical white-dwarf luminosity functions for two phase diagrams of the carbon-oxygen dense plasma. *Astron. Astrophys.* **193**, 141-147 (1988).

[6] Segretain, L. *et al*. Cooling theory of crystallized white dwarfs. *Astrophys. J.* **434**, 641-651 (1994).

[7] Althaus, L. G., García-Berro, E., Isern, J., Córsico, A. H. & Miller Bertolami, M. M. New phase diagrams for dense carbon-oxygen mixtures and white dwarf evolution. *Astron. Astrophys.* **537**, A33 (2012).

[8] Horowitz, C. J., Schneider, A. S. & Berry, D. K. Crystallization of Carbon-Oxygen Mixtures in White Dwarf Stars. *Phys. Rev. Lett.* **104**, 231101 (2010).

[9] Hughto J. *et al*. Direct molecular dynamics simulation of liquid-solid phase equilibria for a three-component plasma. *Phys. Rev. E.* **86**, 066413 (2012).

[10] Winget et al. An independent method for determining the age of the universe. *Astrophys. J. Lett.* **315**, L77-L81 (1987).

[11] Fontaine, G., Brassard, P. & Bergerson, P. The Potential of White Dwarf Cosmochronology. *Publ. Astron. Soc. Pac.* **113**, 409-435 (2001).

[12] Obertas, A. *et al*. The onset of convective coupling and freezing in the white dwarfs of 47 Tucanae. *Mon. Not. R. Astron. Soc.* **474**, 677-682 (2018).

[13] Garcia-Berro, E. *et al.* A white dwarf cooling age of 8 Gyr for NGC 6791 from physical separation processes. *Nature*. **465**, 194-196 (2010).

[14] Bédard, A., Bergeron, P. & Fontaine, G. Measurements of Physical Parameters of White Dwarfs: A Test of the Mass-Radius Relation. *Astrophys. J.* **848**, 11 (2017).

[15] Gaia Collaboration *et al.* Gaia Data Release 2: Observational Hertzsprung-Russell diagrams. *Astron. Astrophys.* **616**, A10 (2018).

[16] Hansen, B. M. S. *et al.* The White Dwarf Cooling Sequence of the Globular Cluster Messier 4. *Astrophys. J. Lett.* **574**, L155-L158 (2002).

[17] Tremblay, P.-E. Kalirai, J. S., Soderblom, D. R., Cignoni, M. & Cummings, J. White Dwarf Cosmochronology in the Solar Neighborhood. *Astrophys. J.* **791**, 92 (2014).

[18] Gentile Fusillo *et al.* The Gaia Data Release 2 catalogue of white dwarfs and a comparison with SDSS. Submitted to *Mon. Not. R. Astron. Soc.* Preprint at https://arxiv.org/abs/1807.03315 (2018).

[19] Tremblay, P.-E., Ludwig, H.-G., Steffen, M. & Freytag, B. Spectroscopic analysis of DA white dwarfs with 3D model atmospheres. *Astron. Astrophys.* **559**, A104 (2013).

[20] El-Badry, K., Rix, H.-W. & Weisz, D.R. An Empirical Measurement of the Initial-Final Mass Relation with Gaia White Dwarfs. *Astrophys. J. Lett.* **860**, L17 (2018).

[21] Chandrasekhar, S. The highly collapsed configurations of a stellar mass (Second paper). *Mon. Not. R. Astron. Soc.* **95**, 207-225 (1935).

[22] Kleinman, S. J. *et al.* SDSS DR7 White Dwarf Catalog. *Astrophys. J. Supp.* **204**, 5 (2013).

[23] Bergeron, P., Saffer, R.A. & Liebert, J. A spectroscopic determination of the mass distribution of DA white dwarfs. *Astrophys. J.* **394**, 228-247 (1992).

[24] Kalirai, J. S., Richer, H. B., Hansen, B. M. S., Reitzel, D. & Rich, R. M. The Dearth of Massive, Helium-rich White Dwarfs in Young Open Star Clusters. *Astrophys. J. Lett.* **618**, L129-L132 (2005).

[25] Tremblay, P.-E. et al. On the Evolution of Magnetic White Dwarfs. *Astrophys. J.* **812**, 19 (2015).

[26] Kalirai, J. S. *et al.* Ultra-Deep Hubble Space Telescope Imaging of the Small Magellanic Cloud: The Initial Mass Function of Stars with M < 1 Msun. *Astrophys. J.* **763**, 110 (2013).

[27] Bertelli, G., Nasi, E., Girardi, L. & Marigo, P. Scaled solar tracks and isochrones in a large region of the Z-Y plane. II. From 2.5 to 20 Msun stars. *Astron. Astrophys*. **508**, 355-369 (2009).

[28] Potekhin, A. Y. & Chabrier, G. Equation of state of fully ionized electron-ion plasmas. II. Extension to relativistic densities and to the solid phase. *Phys. Rev. E.* **62**, 8554-8563 (2000).

[29] Marigo, P. Chemical Yields from Low- and Intermediate-mass Stars: Model Predictions and Basic Observational Constraints. *Astron. Astrophys*. **370**, 194-217 (2001).

[30] Mestel, L. & Ruderman, M. A. The energy content of a white dwarf and its rate of cooling. *Mon. Not. R. Astron. Soc*. **136**, 27-38 (1967).

Acknowledgements

The research leading to these results has received funding from the European Research Council under the European Union's Horizon 2020 research and innovation programme n. 677706 (WD3D) and under the European Union's Seventh Framework Programme (FP/2007- 2013) / ERC Grant Agreement n. 320964 (WDTracer). This work has made use of data from the European Space Agency (ESA) mission Gaia (https://www.cosmos.esa.int/gaia), processed by the Gaia Data Processing and Analysis Consortium (DPAC, https://www.cosmos.esa.int/web/gaia/dpac/consortium). Funding for the DPAC has been provided by national institutions, in particular the institutions participating in the Gaia Multilateral Agreement. Support for J.J.H. was provided by NASA through Hubble Fellowship grant #HST-HF2-51357.001-A, awarded by the Space Telescope Science Institute, which is operated by the Association of Universities for Research in Astronomy, Incorporated, under NASA contract NAS5-26555.

Author Contributions

P.-E.T. and B.D. have identified and characterised the empirical crystallisation sequence. G.F. made the evolutionary white dwarf models used in this work. N.P.G.F., M.H., and T.C. constructed the *Gaia* white dwarf sample employed in this study and performed the cross-match with other photometric and spectroscopic surveys. P.-E.T., B.G., T.M., J.J.H. and G.F. wrote the text and participated in the elaboration of the argument for a crystallisation sequence. E.C. and T.C. characterised the accuracy of *Gaia* measurements and derived parameters for white dwarfs.

Author Information

Reprints and permissions information is available at www.nature.com/reprints. The authors declare no competing financial interests. Correspondence and requests for materials should be addressed to P.-E.T. (P-E.Tremblay@warwick.ac.uk).

Data Availability Statement

The Gaia DR2 catalogue of white dwarfs used in this study is available from the University of Warwick astronomy catalogues repository, https://warwick.ac.uk/fac/sci/physics/research/astro/research/catalogues/gaia_dr2_white_dwarf_candidates_v2.csv. All modelling was performed with our extensive white dwarf evolution code. We have opted not to make this multi-purpose code available, but the cooling sequences calculated for this work are available upon request to the authors.

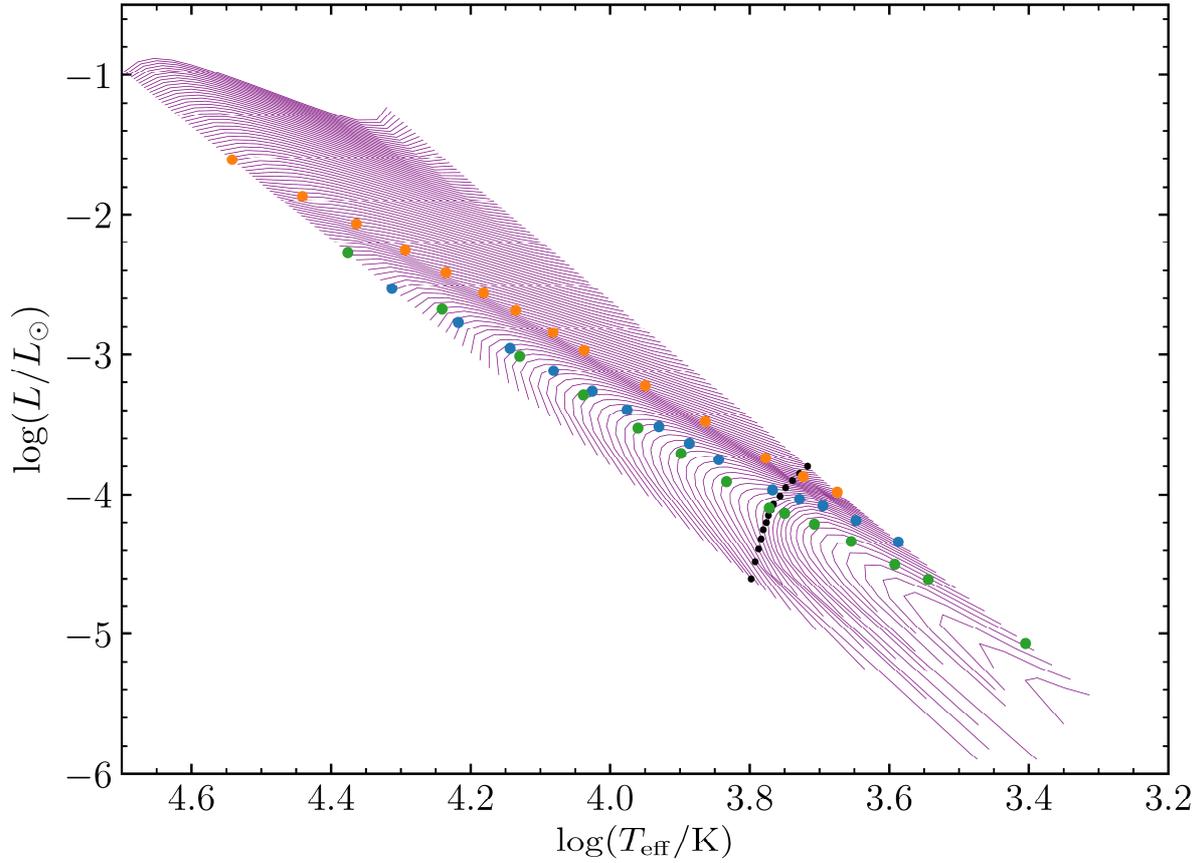

Fig. 1 | **Illustration of the effects of crystallisation on the cooling of white dwarfs**. The closely spaced isochrones in effective temperature-luminosity space connect white dwarfs of the same age (*log* t$_{cool}$ = 7.5 [yr] at the top, with subsequent increments of Δ*log* t$_{cool}$ = 0.02) but with different masses (from 0.4 M$_\odot$ on the low-T$_{eff}$ sides of the isochrones to 1.3 M$_\odot$ on the high-T$_{eff}$ sides). The (variable) density of these many isochrones indicates graphically phases of slowing down and of accelerated cooling. All models are for standard pure-hydrogen atmosphere DA white dwarfs with the same envelope stratification (M$_H$/M$_{WD}$ = 10$^{-4}$ and M$_{He}$/M$_{WD}$ = 10$^{-2}$) and core composition ($^{12}$C/$^{16}$O in 50/50 proportions by mass fraction, and homogeneously distributed)[11]. The models include the release of latent heat, but no additional energy source associated with phase separation[5-7]. From the top, the series of orange dots indicates the onset of crystallisation at the centre of the

evolving model in selected evolutionary sequences. At that point, as the crystallisation front progresses upward in the star from the centre, latent heat is liberated, forming a crest of isochrones taking the shape of a "transverse" sequence. Since the internal energy is discontinuous between the liquid and solid phases, this is a predicted phase transition of the first order[3]. The series of blue dots indicates the location where 80% of the mass has solidified. Following this event, the most significant effect of crystallisation on the cooling of white dwarfs is the so-called Debye cooling phase[6,11], i.e., the transition, in the solid state, from the classical regime to the quantum regime, indicated through a series of green dots. Finally, the onset of the coupling between the upper convection zone with the degenerate core[25] is illustrated by black dots.

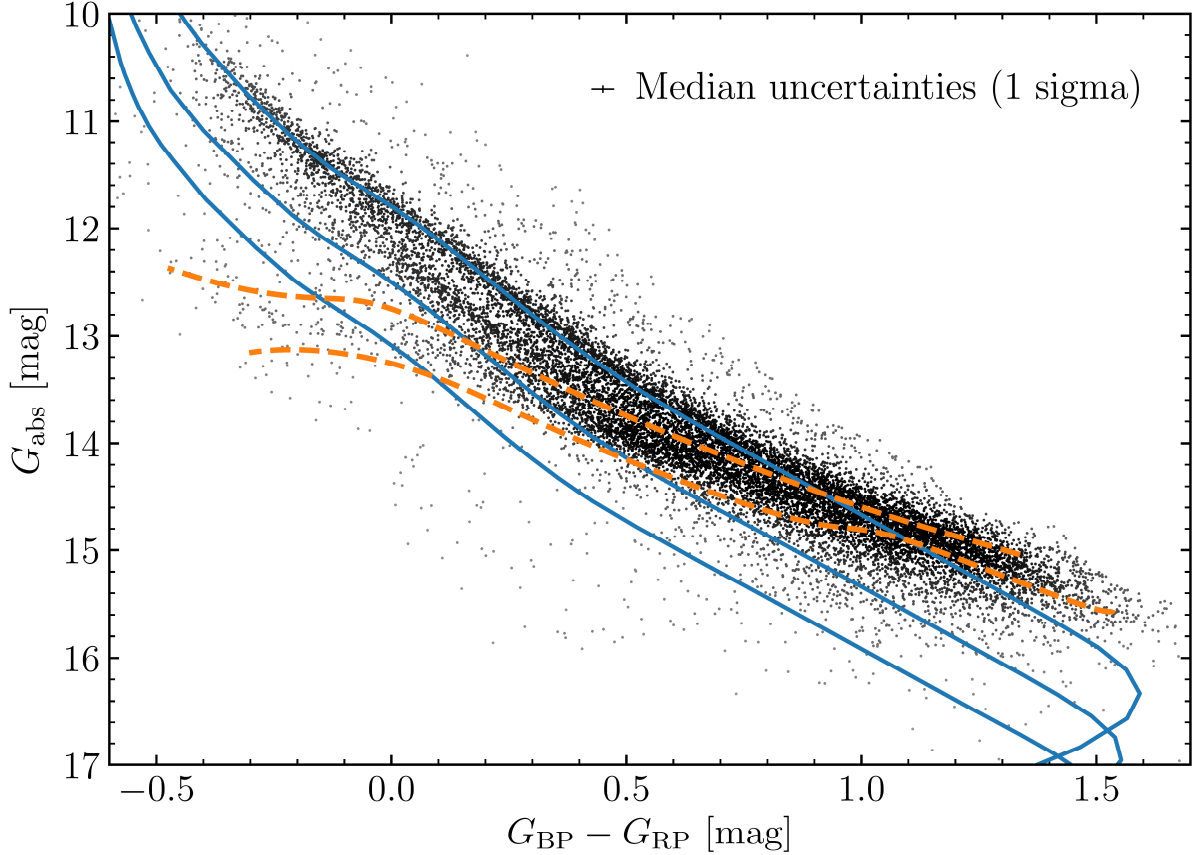

Fig. 2 | **Observational *Gaia* colour-magnitude H-R diagram for white dwarfs within 100 pc**. Dereddened $G$, $G_{BP}$, and $G_{RP}$ photometry along with parallaxes are used for 15,109 white dwarf candidates with *Gaia* data reliable enough to derive atmospheric parameters[18]. For visualisation purposes, the data are shown in a greyscale according to a Gaussian kernel density estimate, and with a power-law scaling of exponent 0.25. Two orange dashed sequences illustrate where evolutionary models predict that 20% (top sequence) and 80% (bottom sequence) of the total white dwarf mass has crystallised. The higher density of white dwarfs within that region corresponds to the transverse sequence discussed in the text. Three evolutionary models at 0.6, 0.9, and 1.1 $M_\odot$ from the top to bottom (blue solid lines) illustrate the evolution of H-atmosphere white dwarfs with thick hydrogen layers[11]. The bifurcation of the observed cooling sequence in two separate

tracks in the range −0.1 < $G_{BP} - G_{RP}$ < 0.6, and above the orange-dashed curves, is not caused by crystallisation but has been interpreted as the different positions of hydrogen- and helium-atmosphere white dwarfs[18,20].

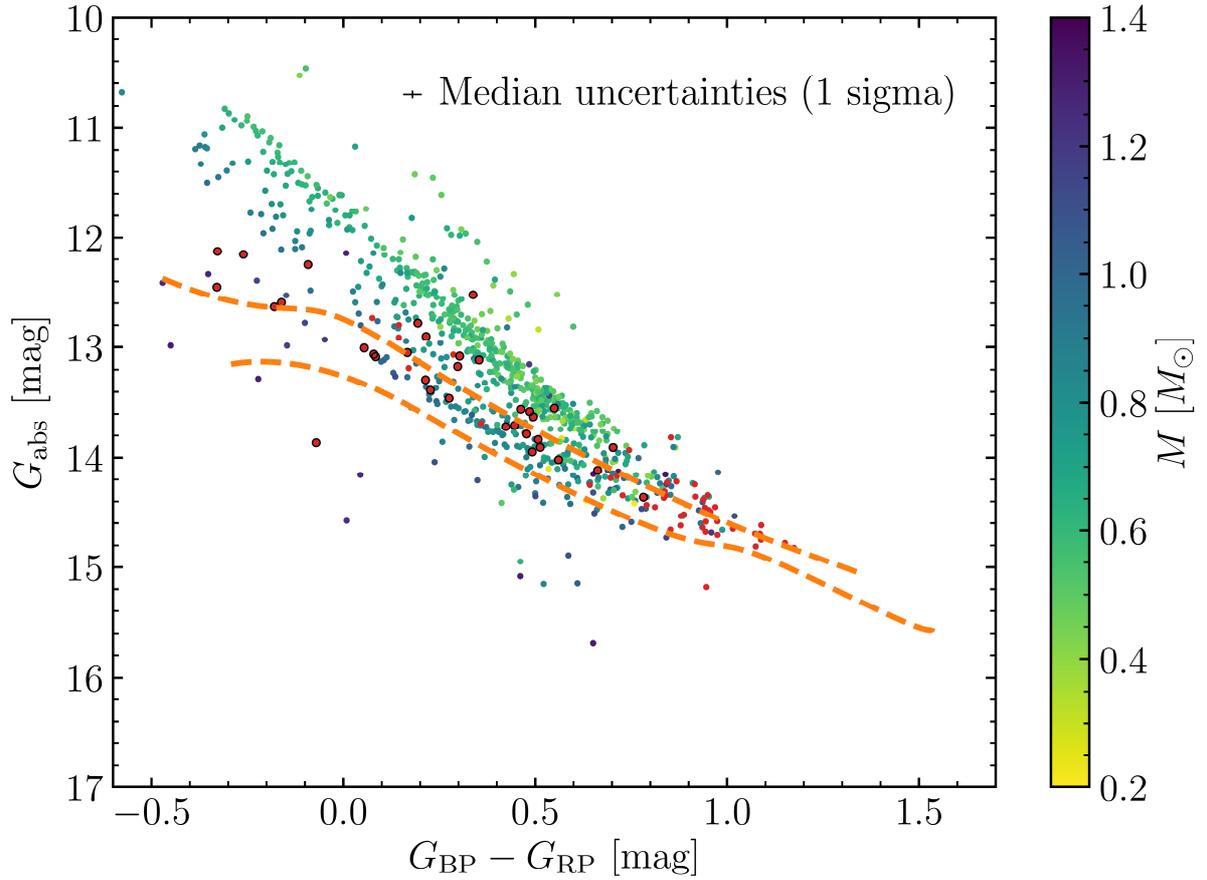

Fig. 3 | **Observational *Gaia* H-R diagram for white dwarfs with SDSS spectra**. Included are 798 objects within 100 pc that show the presence of hydrogen Balmer lines and no helium lines or red excess from a companion[18]. White dwarfs are colour coded (see right-hand scale) for their independent spectroscopic masses[19,22,23] except when lines are too weak to derive masses ($\sigma_M/M > 50\%$, red dots), or there is evidence of a magnetic field (>2 MG) from Zeeman line splitting (red dots with black outlines). Two orange dashed sequences illustrate where evolutionary models predict that 20% (top) and 80% (bottom) of the total white dwarf mass has solidified. This region where the bulk of the crystallisation occurs shows an overdensity of objects.

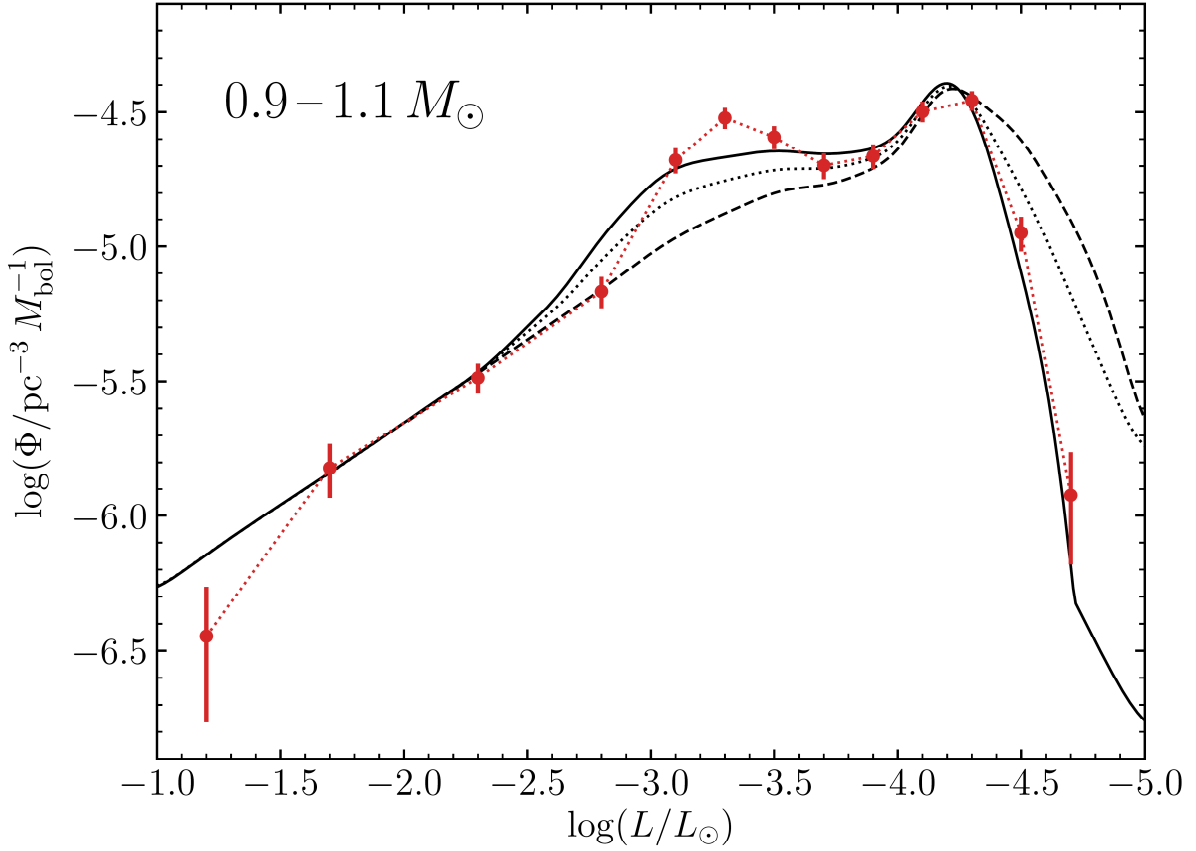

Fig. 4 | **Luminosity function for massive white dwarfs within 100 pc**. Stellar remnants with *Gaia* derived masses between 0.9 and 1.1 $M_\odot$ were used to calculate the observed luminosity function (connected red dots). Error bars are from number statistics (1 sigma). The first peak on the left is a direct observational signature of crystallisation in white dwarfs. The second peak on the right followed by a sharp drop off at smaller luminosities is caused by the finite age of the Galactic disk[10]. Three different predicted luminosity functions are employed to illustrate the physics of crystallisation. All models use the same assumptions on Galactic evolution, including an age of 10 Gyr for the disk. In the standard case (solid line) both the latent heat released from crystallisation and the gravitational energy released from $^{16}$O sedimentation are included. The dotted curve neglects phase separation while still including the release of latent heat, while the

dashed curve neglects both latent heat and phase separation. In the latter case the equation-of-state still transits from liquid to solid as otherwise the solution is not physical. The three models are arbitrarily normalised based on the 2nd and 3rd highest luminosity bins.